\documentclass[twocolumn]{aastex631}

\usepackage{newtxtext,newtxmath}
\usepackage[T1]{fontenc}
\usepackage{ae,aecompl}

\usepackage{graphicx}
\usepackage{amsmath}
\usepackage{amssymb}

\usepackage[caption=false]{subfig}

\usepackage[normalem]{ulem}
\usepackage{ragged2e}
\usepackage{xcolor}
\usepackage{color}

\begin{document}

\title{AGN Feedback-Induced Stellar Density Expansion in the Inner Regions of Early-Type Galaxies}

\correspondingauthor{Fazeel Khan}
%\email{fmk5060@nyu.edu, khanfazeel.ist@gmail.com}

\author[0000-0002-5707-4268]{Fazeel Mahmood Khan}
\affiliation{New York University Abu Dhabi, PO Box 129188, Abu Dhabi, United Arab Emirates}
\affiliation{Center for Astrophysics and Space Science (CASS), New York University Abu Dhabi}

\author{\'Angel Rodr\'iguez}
\affiliation{New York University Abu Dhabi, PO Box 129188, Abu Dhabi, United Arab Emirates}
\affiliation{Center for Astrophysics and Space Science (CASS), New York University Abu Dhabi}

\author[0000-0002-8171-6507]{Andrea V. Macci\`o}
\affiliation{New York University Abu Dhabi, PO Box 129188, Abu Dhabi, United Arab Emirates}
\affiliation{Center for Astrophysics and Space Science (CASS), New York University Abu Dhabi}

\author{Smarika Sharma}
\affiliation{New York University Abu Dhabi, PO Box 129188, Abu Dhabi, United Arab Emirates}
\affiliation{Center for Astrophysics and Space Science (CASS), New York University Abu Dhabi}

\begin{abstract}

 Observations indicate that early-type galaxies exhibit varying slopes in the relation between their central stellar surface density and stellar mass ($\Sigma_1 - M_{\star}$). Low-mass galaxies tend to follow a steep slope, close to one, while the slope flattens for high-mass early type galaxies. In our study, we investigate the $\Sigma_1 - M_{\star}$ scaling relation and its evolution using the NIHAO suite of cosmological simulations and compare our findings with recent results from the MaNGA survey. Our analysis shows that NIHAO galaxies successfully reproduce the observed scaling relation based on MANGA survey. Our analysis suggests that AGN feedback plays a critical role in flattening the $\Sigma_1$ slope by expelling gas from galactic centers, leading to a decrease in both stellar and dark matter density as the gravitational potential becomes shallower. To further support our findings, we conducted high-resolution N-body simulations, which confirmed that 
  ({\it sudden}) gas removal does substantially alter the stellar density in the central region, consistent with results from NIHAO. Furthermore our numerical experiments show that even if  the same amount of gas is re-accreted on a typical ({\it longer}) free-fall time, it is not able to restore the original stellar density. Our study concludes that AGN feedback assisted gas removal presents a plausible explanation for decline in central stellar surface density as observed in massive elliptical galaxies.

  %changes in the stellar density profile within 1 kpc are consistent with those observed in the NIHAO simulations. The re-introduction of gas potential over a period of 500 Myr restores some of the central density but falls well short of the original value. Our study concludes that AGN feedback assisted gas removal presents a plausible explanation for decline in central surface density as observed in massive elliptical galaxies.

\end{abstract}

%% Keywords should appear after the \end{abstract} command. 
%% The AAS Journals now uses Unified Astronomy Thesaurus concepts:
%% https://astrothesaurus.org
%% You will be asked to selected these concepts during the submission process
%% but this old "keyword" functionality is maintained in case authors want
%% to include these concepts in their preprints.
\keywords{
\href{http://astrothesaurus.org/uat/16}{Active galactic nuclei (16)};
\href{http://astrothesaurus.org/uat/159}{Black hole physics (159)};
\href{http://astrothesaurus.org/uat/573}{Galaxies (573)};
\href{http://astrothesaurus.org/uat/594}{Galaxy evolution (594)};
\href{http://astrothesaurus.org/uat/612}{Galaxy physics (612)};
\href{http://astrothesaurus.org/uat/615}{Galaxy properties (615)};
%\href{http://astrothesaurus.org/uat/1914}{Regression (1914)};
%\href{http://astrothesaurus.org/uat/2031}{Scaling relations (2031)};
\href{http://astrothesaurus.org/uat/1663}{Supermassive black holes (1663)}
}
%%%%%%%%%%%%%%%%%%%%%%%%%%%%%%%%%%%%%%%%%%%%%%%%
	
	%%%%%%%%%%%%%%%%% BODY OF PAPER %%%%%%%%%%%%%%%%%%

\section{Introduction}\label{sec-intro}

%\subsection{Massive black holes}

Galaxies grow in size and mass through mergers and gas accretion. The accreted gas can reach the galaxy's center, where it contributes to an increase in stellar mass and density by forming new stars. This gas can also trigger AGN (Active Galactic Nucleus) activity by matter falling onto central supermassive black hole, a process which can release enormous amount of energy comparable to $10 \%$ of the rest mass energy, significantly heating the gas and expelling it from the central part. Feedback from AGN is thought to regulate the growth of the supermassive black hole (SMBH) and star formation within the galaxy. This regulation is evident from several strong correlations between SMBHs and their host galaxies, such as the SMBH mass versus velocity dispersion \citep{ferrarese+00} and the SMBH mass versus galaxy or bulge mass \citep{har04,Kormendy+13,mcconnell+13,sav13}.  Moreover, various galaxy properties are interrelated; for instance, the star formation main sequence correlates a galaxy's mass with its star formation rate \citep{elbaz07,Noeske07,elbaz11,whitaker12}, while the size-stellar mass relation \citep{tru20} and the relation between the central surface density within one kpc and stellar mass ($\Sigma_1$-$M_\star$) \citep{Fang2013,barro2017,chen20}, highlight additional key connections between galaxy characteristics. These correlation are key in our understanding of merger histories, gas accretion, star formation, feedback and quenching of galaxies.

 As galaxies evolve, their central surface and volume densities increase due to various physical processes. Gas rich galaxy mergers, disk instabilities and bar related instabilities drive gas towards the center causing star formation and stellar density growth in the center.
The central stellar surface density and volume density of galaxies provide important insights about specific star formation rate and galaxy quenching\citep{Brinchmann2004,kauffmann:2006,maier:2009}. \citet{Cheung2012} demonstrated that $\Sigma_1$ correlates strongly with galaxy color, highlighting the inner mass distribution as a key indicator of quenching.
\citet{Fang2013} studied $\Sigma_1$-$M_\star$ relation and divided the galaxies in blue sequence, green valley and red sequence. It was observed that red sequence galaxies have higher values of $\Sigma_1$ for a given $M_\star$ than those of blue sequence. This suggests that star formation quenching is accompanied by an increase in $\Sigma_1$. \citet{barro2017} investigated $\Sigma_1$-$M_\star$ and noticed clear distinction between star forming and quiescent galaxies and concluded that a dense core is a pre-requisite for quenching of star formation. Their study suggests that, in addition to a dense core, morphological transformation from an exponential (disk-like) profile to a more concentrated Sérsic profile (with $n \geq 2$) is also necessary for quenching, underscoring the importance of a spheroidal component in the quenching process.

\citet{Arora21} examined the structural properties of 4585 galaxies from optical and mid-infrared data extracted from the Dark Energy Sky Instrument Legacy Imaging Survey (DESI) and the Wide-Field Infrared Survey Explorer (WISE) survey. They find that $\Sigma_1$-$M_\star$ relation can be fitted by piece-wise linear fits with two different slopes. For lower-mass galaxies ($\log M_\star \leq 10.73$), the $\Sigma_1 - M_\star$ relation is nearly linear, with a positive slope close to 1. This suggests that the inner and outer regions of these galaxies co-evolve, driven by star formation and environmental interactions. For Early Type Galaxies (ETGs) with masses $\log M_\star \geq 10.73$ stellar masses, the slope drastically changes to 0.12 suggesting that for these massive galaxies, there is likely very little star formation but an ongoing accretion can cause very gradual increase in central mass density. It is not well understood why the central density of massive ETGs saturates as they approach towards high mass end. 

One suggested mechanism is related to core scouring by SMBH binaries, as they deposit orbital energy into the surroundings \citep{graham+04,mer06,khan+12a}
  As almost all massive galaxies host a central SMBH, a major merger between them would inevitably result in the formation of a pair of SMBHs in the product galaxy \citep{Begelman:1980,Callegari2011,2016ApJ...828...73K}. During the course of their coalescence, an SMBH binary can remove the equivalent of a few times its mass in stellar mass as it exchanges its orbital energy with surrounding stars \citep{mer06, khan+12a}. This mechanism is thought to be responsible for the formation of the core in massive ellipticals \citep{graham+04,rantala2018,Nasim2021}. Moreover, as the two SMBHs merge, the asymmetric emission of gravitational waves causes the merged SMBH to recoil, and it has been shown that this recoil can further contribute to increase the core size \citep{2008ApJ...678..780G,Khonji:2024}. 

%A suggested mechanism in the literature that can be relevant here is that of core scouring by SMBH binaries as they deposit orbital energy to the surroundings \citep{graham+04,mer06,khan+12a}. This effect can create cores in massive ellipticals that can extend up to a kilo parsec.

In this study, we propose another mechanism that can cause the mass depletion in the center of massive galaxies. Gas expulsion from the centers of massive galaxies by the AGN feedback and the subsequent dynamical expansion of stars and dark matter can provide a plausible explanation for the observed flattening of the $\Sigma_1 - M_\star$ relation. 
Cosmological simulations are an exquisite tool to understand the formation and evolution of galaxies as they grow across the cosmic landscape. To understand the physical mechanisms that can lead to a decrease in central stellar density, we conducted an extensive analysis of how this density changes in zoom-in cosmological simulations. Our sample includes galaxies of various stellar masses from the Numerical Investigation of Hundred Astrophysical Objects (NIHAO) simulation suite \citep{Wang2014,Blank2019}. This approach helps us explore the physical processes that govern the $\Sigma_1 - M_\star$ relation, particularly for the most massive elliptical galaxies.

As is the case with cosmological simulations in general, NIHAO simulations have limited spatial and mass resolution. In order to qualitatively validate the conclusions reached on the basis of this study, we also performed controlled experiments with high-resolution $N-$body simulations without gas physics. 

The paper is organized as follows. Section 2 provides a detailed analysis of the NIHAO simulations, focusing on the time evolution of key physical parameters. In Section 3, we describe the properties of our selected galaxies for which we perform direct N-body simulations. Section 4 describes our code, and Section 5 elaborates the results of our N-body experiments. Finally, Section 6 summarizes and concludes the study.

\section{Cosmological Zoom-in Simulations NIHAO} \label{NIHAO}

%%-------------------------------------------------------------------------%
\begin{figure*}
\centering{
  \resizebox{0.95\hsize}{!}{\includegraphics{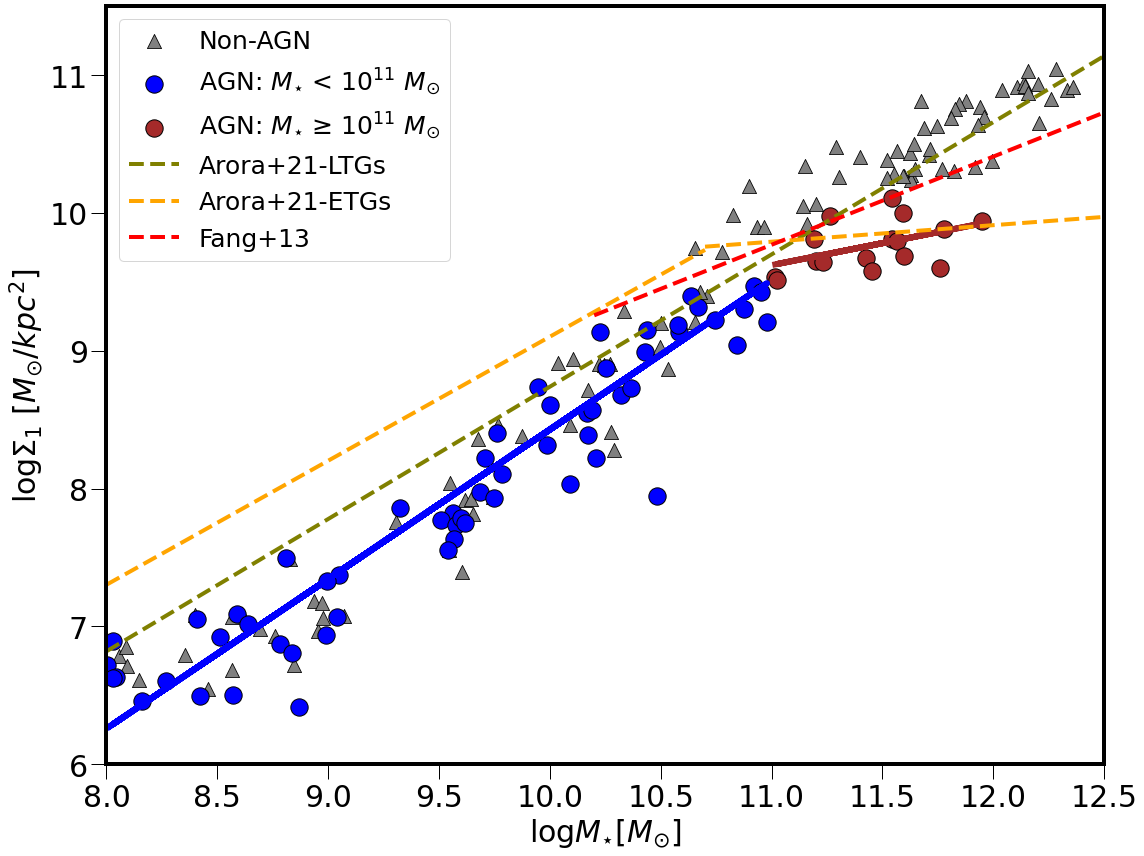}}
  }
\caption{
$log\Sigma_{1}$ vs $logM_{\star}$ plot for all NIHAO galaxies both with (filled circles) and without (filled triangles) central SMBH and AGN feedback at redshift 0. The $\Sigma_{1}$ is measured for stars inside the sphere of radius 1 kiloparsec from the halo center.  Blue and brown full lines are fit to NIHAO AGN sample galaxies below and above the critical mass $M_{\star}$ = $10^{11}$ $M_{\odot}$, respectively.
The green and orange dashed lines are fits of $\Sigma_{1}$ vs $M_{\star}$ from \citet{Arora21} for early type and late type galaxies respectively. The red dashed line is fit for massive red sequence galaxies in the sample of \citet{Fang2013}.}  \label{fig:allgal}
\end{figure*} 
%------------------------------------------------------------------------

\begin{figure*}
    \centering{
    \resizebox{0.99\hsize}{!}{\includegraphics[angle=0]{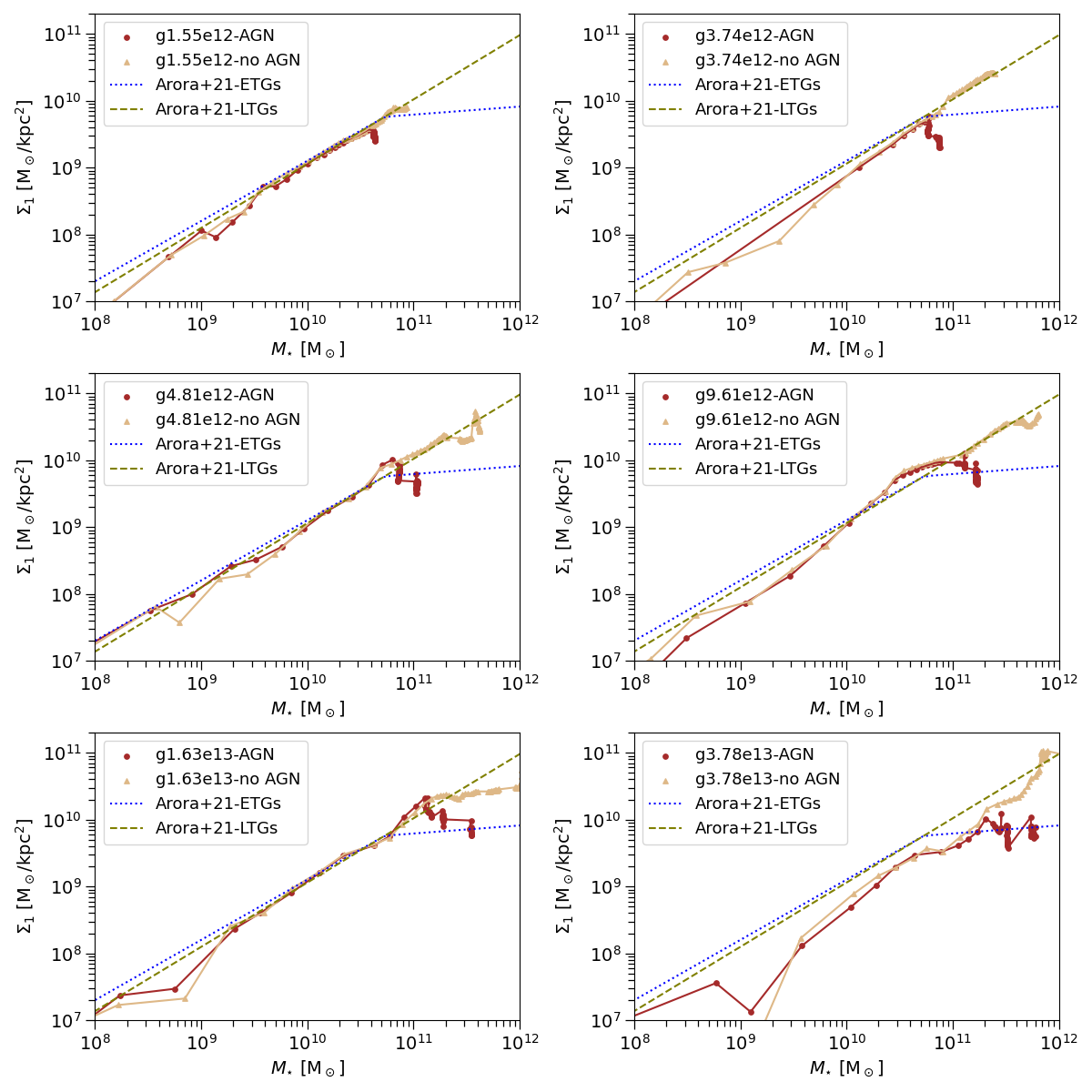}}
    }
    \caption{$\Sigma_{1}$-$M_{\star}$ for a sample of 6 NIHAO galaxies as they evolve and grow in size over cosmic history. The data is plotted for each galaxy for simulations with (brown circles) and without central black hole feedback (orange triangles). The dashed and dotted lines in each plot are observed $\Sigma_{\star}$-$M_{\star}$ relation of \citet{Arora21} for early and late type galaxies respectively.}
    
\end{figure*}\label{fig:nihaosigmavsmstar}

The NIHAO suite is a very large set of cosmological zoom-in hydrodynamical simulations performed using the {\sc gasoline2} code \citep{Wadsley2017}, featuring high resolutions of $10^{6}$ particles per halo and encompassing galaxies with total masses ranging from $10^{8}$ to $10^{13}$ $M_{\odot}$. 
NIHAO simulations have successfully reproduced key scaling relations that are fundamental to galaxy evolution, such as the $M_{\star}$–halo mass relation \citep{Wang2017}, Tully-Fisher \citep{Dutton2015} and relations between $M_{\star}$, the mass of the central supermassive black hole and the star formation rate \citep{Blank2019, Blank2021}.

For this study we use two sets of NIHAO simulation, with and without the implementation of supermassive black hole accretion and feedback, as detailed in \cite{Blank2019}; we will refer to these two distinct sets of simulations as AGN and no AGN respectively.

\subsection{$\Sigma_1 - M_\star$ relation}

The set of NIHAO galaxies allows the study of galaxy evolution at various times during the history of their evolution. We use publicly available package \textit{pynbody} \citep{pon13} to extract physical parameters like stellar mass and stellar surface density inside 1 kpc.

%There are two series of NIHAO simulations of galaxies with and without a central Supermassive Black Hole (SMBH) and AGN feedback. 
In Figure \ref{fig:allgal}, we plot the average stellar surface density inside 1 kpc against the galaxy mass at redshift zero for smaple of our NIHAO galaxies in both series. For reference, we also over-plot the observed correlations for early-type (ETGs) and late-type galaxies (LTGs) as reported by \citet{Arora21} and \citet{Fang2013}. We notice that galaxies below $M_{\star} \leq 10^{10}$ $M_{\odot}$ are predominantly star forming LTGs. In the mass range $10^{10} \geq  M_{\star} \leq 10^{11}$ NIHAO galaxies exhibit mixed morphologies, transiting from LTGs to ETGs where as $M_{\star} \geq 10^{11} M_{\odot}$ are elliptical with very little star formation. We employ a piecewise linear fit for low- and high-mass galaxies in our sample differentiating at a mass of $M_{\star} \geq 10^{11} M_{\odot}$. For our combined sample of LTGs and lower mass ETGs, we obtain a slope of $1.08$ very close to unity. On the other hand high mass ETGs fit has a slope of $0.32$. This value is situated near the midpoint of the range defined by the results of \citet{Arora21}, with reported a value of $0.12$, and \citet{Fang2013}, who reported a value of $0.66$ for their sample of high mass ETGs.

The low stellar mass galaxies in both sets of NIHAO simulations follow the observed behavior of a linear fit with slope close to unity for LTGs and ETGs. However, for the more massive galaxies, only the ones with AGN feedback follow the observed shallow correlation while the ones with no AGN feedback continue to have an unchanged steep slope. This suggests that the black hole accretion and feedback mechanisms are key to produce the change in the slope for the the observed $\Sigma_1 - M_\star$ relation.

Cosmological simulations offer the unique advantage of allowing us to track a galaxy's evolution across cosmic time, unlike direct observations. To examine how $\Sigma_1 - M_\star$ evolves as a galaxy gains mass, we plot the relation for six massive galaxies in NIHAO sample (namely {\it g1.55e12, g3.74e12, g4.8e12, g9.61e12, g1.63e13, g3,78e13})\footnote{In NIHAO the name of a galaxy represents its total mass in the low resolution Nbody only simulation.}. Figure \ref{fig:nihaosigmavsmstar} contains the same galaxies for two sets of cosmological simulations, with and without the implementation of AGN feedback.

%%-------------------------------------------------------------------------%
%\begin{figure}
%\centerline{
%  \resizebox{0.95\hsize}{!}{\includegraphics{figs/Candidates_Stellar_Surface_Density_vs_Stellar_Mass.png}}
%  }
%\caption{
%bla bla bla} \label{fig:nihaosigmavsmstar}
%\end{figure}
%------------------------------------------------------------------------
\begin{figure*}
    \centering{
%    \resizebox{0.99\hsize}{!}{\includegraphics[angle=0]{figs/Candidates_Stellar_Surface_Density_vs_Time-3.png}}
    \resizebox{0.99\hsize}{!}{\includegraphics[angle=0]{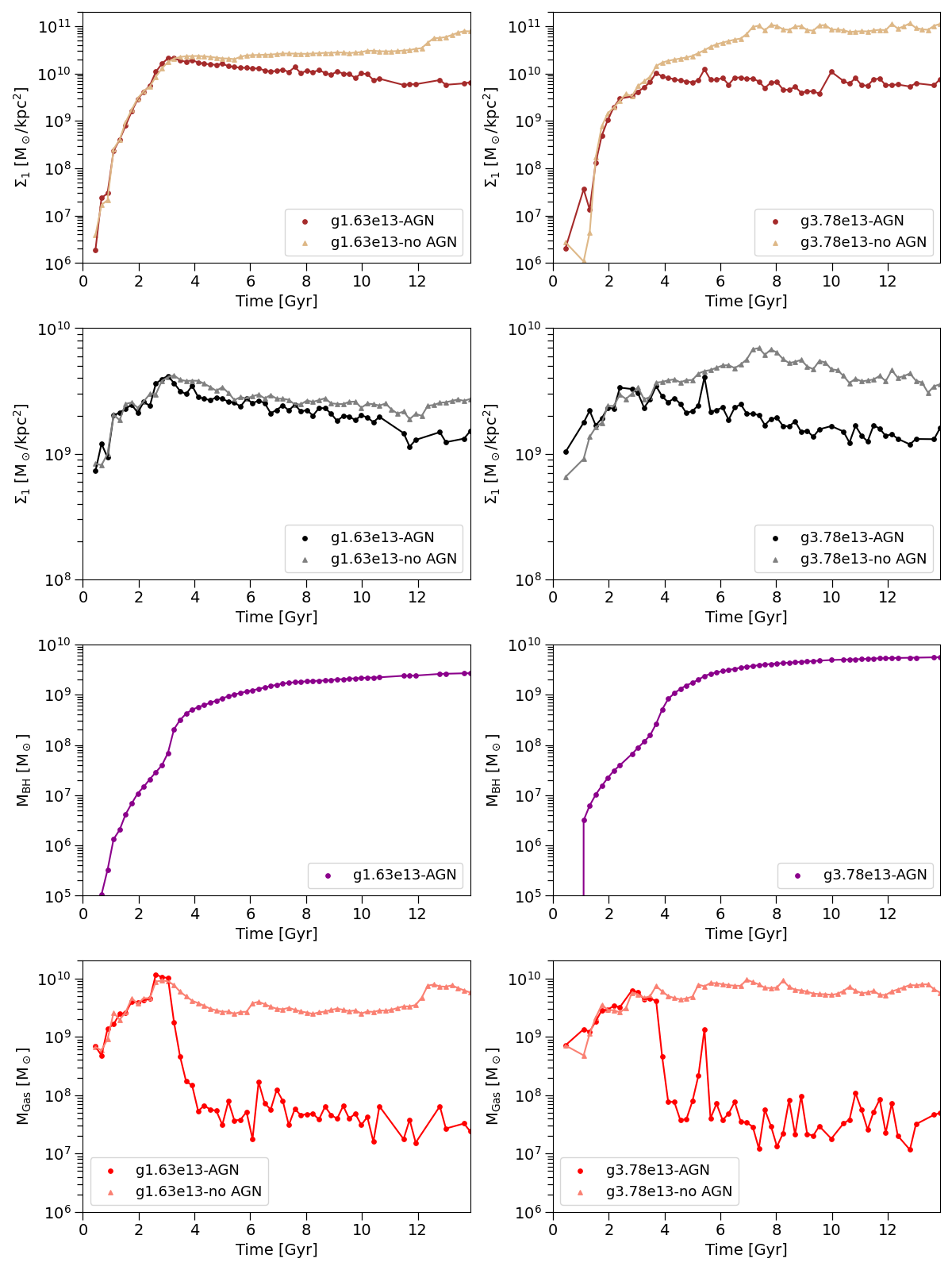}}
    }
    \caption{Time evolution of various physical parameters for simulations with and without AGN feedback for galaxies g1.63e13 (left panels) and g3.78e13 (right panels). Top panel shows $\Sigma_1$ time evolution for stars, second panel shows the same for dark matter, 3rd panel shows the SMBH mass growth as a function of time and finally the bottom panel shows the gas mass inside 1 kpc across entire time evolution.}
    \label{fig:timeevol1}
\end{figure*}
%-------------------------------------------------------------------------

%------------------------------------------------------------------------
\begin{figure*}
    \centering{
%    \resizebox{0.99\hsize}{!}{\includegraphics[angle=0]{figs/Candidates_Stellar_Surface_Density_vs_Time-3.png}}
    \resizebox{0.99\hsize}{!}{\includegraphics[angle=0]{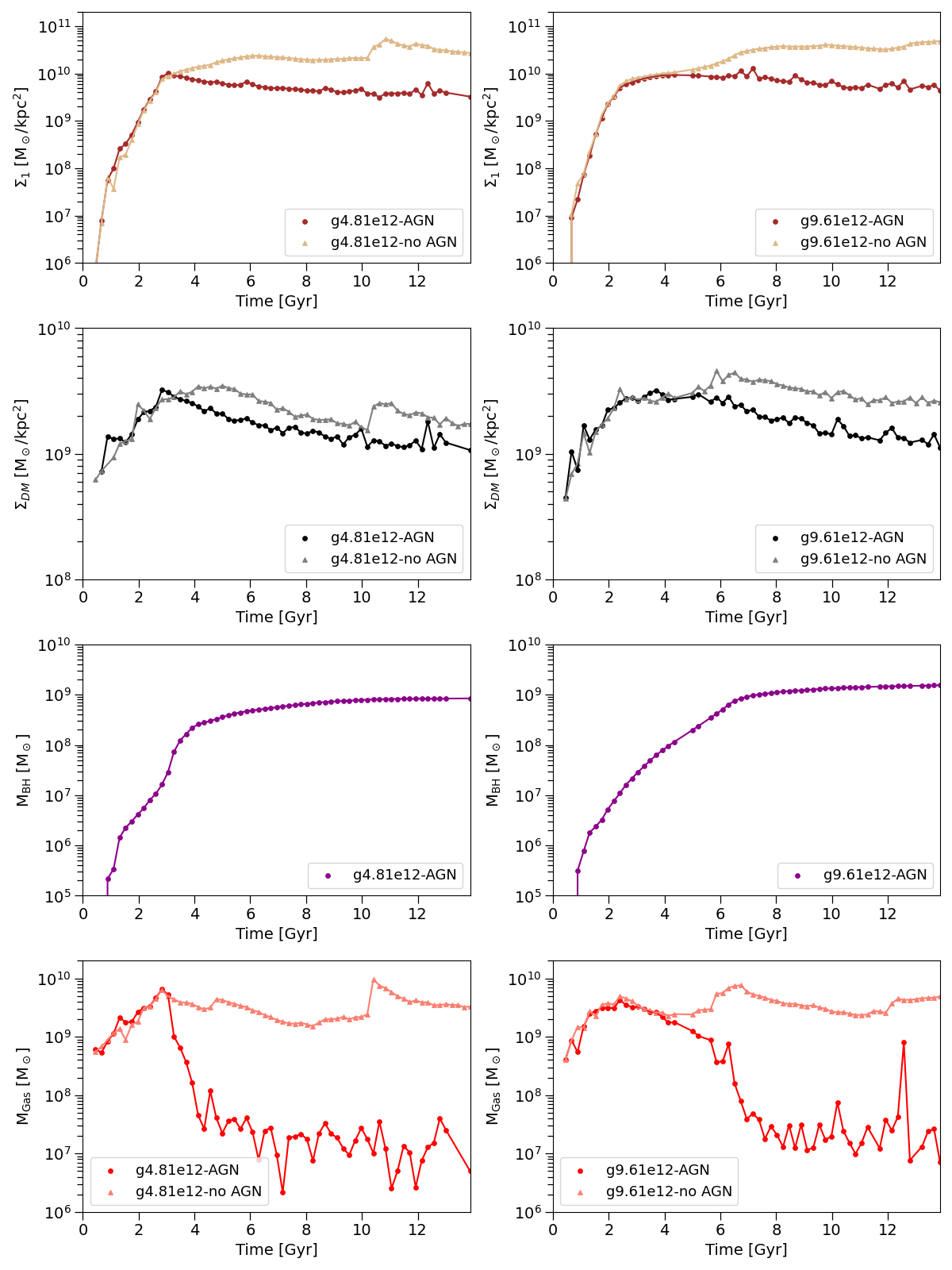}}
    }
    \caption{The same as in figure \ref{fig:timeevol1}, this time for g4.81e12 and g9.61e12. }
    \label{fig:timeevol2}
\end{figure*}
%-------------------------------------------------------------------------

%------------------------------------------------------------------------
\begin{figure*}
    \centering{
%    \resizebox{0.99\hsize}{!}{\includegraphics[angle=0]{figs/Candidates_Stellar_Surface_Density_vs_Time-3.png}}
    \resizebox{0.99\hsize}{!}{\includegraphics[angle=0]{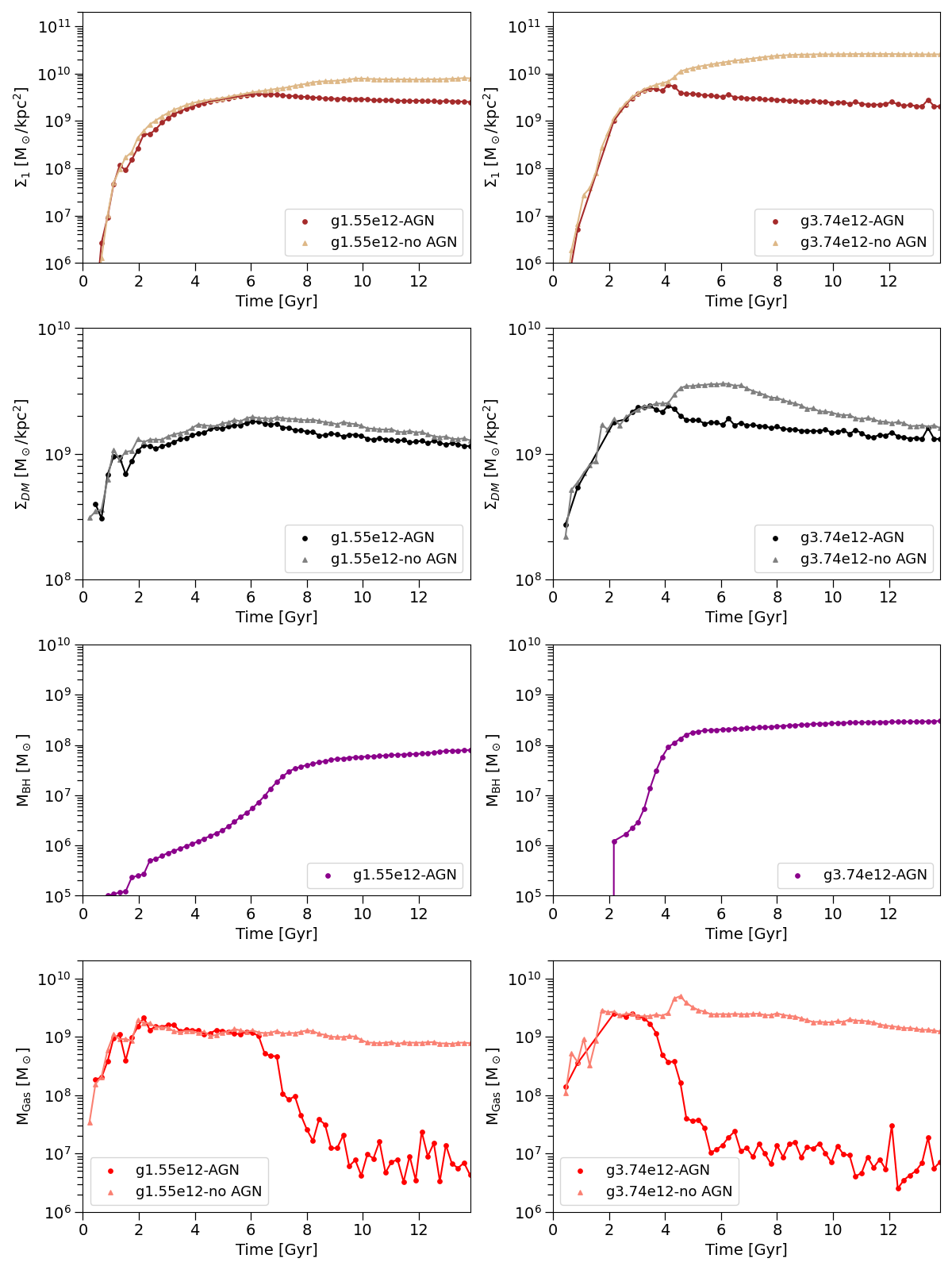}}
    }
    \caption{The same as in figure \ref{fig:timeevol1}, this time for g1.55e12 and g3.74e12.}
    \label{fig:timeevol3}
\end{figure*}
%-------------------------------------------------------------------------

%\begin{figure}
%    \centering
%    \includegraphics[width=0.5\textwidth]{figs/Candidates_Stellar_Surface_Density_vs_Stellar_Mass-3.png}
%    \caption{Evolution of $\Sigma_{\star}$-$M_{\star}$ correlation as galaxies evolve and grow in size over their cosmic history. The data is plotted for each galaxy for simulations with and without central black hole feedback. The dashed line in each plot is the observed $\Sigma_{\star}$-$M_{\star}$ relation of \citet{Arora21}.}
%    \label{fig:nihaosigmavsmstar}
%\end{figure}

We notice that for most of their mass growth histories, galaxies with and without central black holes follow the same trend, except that as their stellar masses grow to about $10^{11}$ $M_{\odot}$, the  $\Sigma_{\star}$-$M_{\star}$ correlation starts to differ for the two set of simulations. The galaxies without AGN feedback keep on growing following the same steep linear relation as observed for low mass galaxies where as those with AGN feedback tend to drop off this correlation and move towards the correlation with much shallower slope as observed in massive galaxies (dotted blue line). 
We also observe sudden drops in the central stellar surface density by almost of a factor of 2 and it seems that these drops are crucial in bringing massive galaxies towards the observed $\Sigma_1 - M_\star$ relation.

\subsection{$\Sigma_1$ - Time}

In figures \ref{fig:timeevol1}, \ref{fig:timeevol2} and \ref{fig:timeevol3}. 
we present the time evolution of various physical quantities in both sets of simulations. 
Top panel of these figures represent the time evolution of the $\Sigma_{\star}(t)$ for the same set of galaxies as shown in figure \ref{fig:nihaosigmavsmstar}.
We notice that galaxies of both simulations initially have an almost identical evolution in a low stellar mass regime. However at around $t=4$ gigayears, corresponding to the galaxies obtaining a stellar mass higher than $10^{10.7}$ M$_{\odot}$, the growth splits up, remaining roughly constant for the case where there is a black hole present in the galaxy. On the other hand, the galaxies without a black hole continues to have an increase in their stellar surface density only reaching a plateau at 10 Gyr. 

A closer look at these plots reveals that galaxies with feedback from central black hole tend to have a negative slope in $\Sigma_1$ during the later phase of evolution, indicating the decrease in stellar surface density as the galaxy continues to evolve. 
%Furthermore, the stellar surface density as a function of time can reveal insight into the merger history of the galaxy. We witness sudden jumps in the $\Sigma_1$ profile due to merger induced injection of matter into the central part of the galaxy. These mergers, however, are not able to change the slope of the $\Sigma_1 - M_*$ relation over long time evolution.

The second panel of these figures shows the time evolution of dark matter surface density inside 1 kpc for the selected subset of galaxies. Again we notice the drop in $\Sigma_1$ happening at about the same time as witnessed for stellar profiles but for dark matter the difference between $\Sigma_1$ for AGN and no AGN cases is less pronounced when compared with $\Sigma_1$ for stars.

\subsection{$M_{\text{BH}}$ - Time} 

We examine the mass growth of the black holes in our galaxy sample as a function of time which is presented in the third row of figures 3-5. We notice that SMBHs grow rapidly during the first 4 Gyrs and then retain roughly the same mass with very little growth during rest of their cosmic evolution history. The black holes grow roughly 3 orders of magnitude very early during the simulation \citep[see][for more details on SMBH accretion in NIHAO]{Blank2019,Soliman2023}. It can be observed that the more massive galaxies host black holes with the fastest growth, while those at lower masses contain black holes that plateau in mass at a later time. For the most massive galaxies, g3.78e13 shown in this sample, SMBH mass growths plateaus at roughly 4 Gyr whereas the same happens much later around 8 Myr for the least massive galaxy g1.55e12 in our sample. Additionally, the SMBH reaches much higher mass in more massive galaxies when compared to those having smaller masses.

\subsection{$M_{\text{gas}}$ - Time}

The evolution of the amount of gas within 1 kpc from the center as a function of time strongly correlates with the strength of the black hole feedback, which in turn is linked to the black hole accretion. Bottom panels of the figures 3-5 show time evolution of the amount of gas inside 1 kpc for our selected set of galaxies in NIHAO simulation. We see a general trend of gas mass growth in the first 4 gigayears of the simulation, corresponding once again to the period of strong gas inflow \citep{Waterval2024}, that leads to a rapid increase of both stellar and balck hole mass.
However, after 4 gigayears, there is a considerable drop in the amount of gas within 1 kpc for all galaxies. This drop in the amount of gas is so large that essentially almost all of the gas is blasted away from the center of the galaxy and gas mass falls by two orders of magnitude. This peak of gas expulsions is related to AGN feedback powered by the fast BH growth, as shown in the third row of figures 3-5.
On the other hand, galaxies without AGN feedback continue to accumulate gas over time and eventually plateau roughly around $10^{10}$ $M_{\odot}$. 
During next 8-10 Gyr of evolution, galaxies with AGN have  marginal gas accretion in the inner region retaining a very low gas fraction in their center. 

Very interestingly this massive gas mass expulsion is highly correlated with the decrease of both stellar and dark matter density within 1 kpc, suggesting an AGN driven expansion of these collision-less components {\it a la} \cite{Pontzen2012}, very similar to the mechanism thought to be responsible of erasing dark matter cusps in dwarf galaxies \citep[e.g.][]{Governato2010, Tollet2016}.

%never manage to get the gas contents of this magnitude again. Most importantly, this decline in gas mass is linked to a decrease in both stellar and dark matter surface density, as well as an end to the rapid SMBH mass growth. 

The strong correlation (in all galaxies) between SMBH growth, AGN feedback, gas expulsion and $\Sigma_1$ drop is very suggestive and might point to the first evidence for stellar expansion in massive galaxies in a similar fashion to what is observed on a very different mass scale for ultra diffuse galaxies \citep[e.g.,]{Dicintio2017,Jiang2019}.  

On the other hand the mass resolution and softening of NIHAO for massive galaxies are on the edge of what is needed to capture these dynamical effects on the scale of about 1 kpc.
%
%
%We find strong evidence linking the drop in central surface density to gas expulsion from the core, driven by AGN feedback. However, given the low mass resolution and resulting limited statistics in the central region, along with the larger gravitational softening used in NIHAO—an inherent feature of most cosmological simulations—we approach these observed effects cautiously, as they may be exaggerated by the softened gravitational potential within the inner kiloparsec. 
%
To address this issue and to strengthen our results, we perform a series of controlled N-body simulations using a selected sample of NIHAO galaxies. We generate initial conditions (see section \ref{sec:runs}) for these galaxy models with higher mass resolution and a larger number of particles. For this experiments we employ the direct $N$-body code $\phi$-GPU \citep{berczik+11,Berczik2013} that allows us to model gravity more accurately and achieve significantly smaller softening lengths in the parsec range.

%\begin{figure*}
%    \centering{
%    \resizebox{0.99\hsize}{!}{\includegraphics[angle=0]%{figs/Gas_Mass_vs_Time.png}}
%    }
%    \caption{Amount of gas in central kiloparsec as a function of time for our selected sample of NIHAO galaxies}
%    \label{fig:mgasvstime}
%\end{figure*}

\section{High Resolution N-body Simulations of Selected NIHAO Galaxies }\label{sec:runs}

%\subsection{Initial NIHAO Models (g3.78e13.00304)}

%%%%%%%%
\begin{table*}
\begin{center}
\vspace{-0.5pt}
\caption{Galaxy Parameters} 
\begin{tabular}{l c c c c c c }
\hline
Galaxy Model & Gas Mass ($M_{\odot}$)  &  Stellar Mass ($M_{\odot}$)  & Dark Matter Mass ($M_{\odot}$)  & $r_0$,$\gamma$ (Gas)  &  $r_0$($R_e$),$\gamma$($n$) (Star) &  $r_0$,$\gamma$ (Dark Matter) \\
\hline
g3.78e13 & $5.95 \times 10^{10}$ & $5.08 \times 10^{10}$ & $9.07 \times 10^{10}$ & $2.0,0.1$ & $2.0,0.6$ & $3.0,0.2$ \\

g9.61e12 & $2.53 \times 10^{10}$ & $5.57 \times 10^{10}$ & $7.85 \times 10^{10}$ & $2.0,0.2$ & $1.0,0.9$ & $2.5,0.1$ \\

g4.81e12  & $2.96 \times 10^{10}$ & $1.45 \times 10^{10}$ & $3.90 \times 10^{10}$ & $3.2,0.1$ & $1.2,0.6$ & $3.0,0.1$ \\
\hline
\end{tabular}\label{tab:simparam}
\vspace{15pt}

\tablecomments{Column~1: NIHAO galaxy model.  Column~2: Gas mass inside 10 kpc. Column~3: Stellar mass inside 10 kpc. Column~4: Dark matter mass inside 10 kpc. Columns~5,6,7: Dehnen/Sersic profile fit parameters for Gas, Star and Dark matter profiles.}
\end{center}
\vspace{15pt}
\end{table*}\label{tab:table1}
%%%%%%%%

We selected 3 NIHAO galaxies of various masses at a time before the expulsion of gas by AGN feedback. Table \ref{tab:simparam} presents some of the key parameters of our selected sample. We extract the positions and velocities of the gas, star and dark matter particles from the NIHAO simulation snapshot. The extracted distributions are then fitted with either a \citet{deh93} (equation \ref{eq:dehn}) or \citet{ser63} (equation \ref{eq:ser}) models.

\begin{equation}
\rho(r) = \frac{(3-\gamma) M}{4\pi} \frac{r_0}{r^\gamma (r + r_0)^{4-\gamma}} \label{eq:dehn}
\end{equation} 

where,
 $\rho(r)$ is the density as a function of radius $r$. $M$ is the total mass of the system. $r_0$ is a scale radius. $\gamma$ is a parameter that describes the inner slope of the density profile, with $0 \leq \gamma < 3$.

\begin{equation}
\rho(r) = \rho_0 \left( \frac{r}{R_e} \right)^{-p} e^{-b(r/R_e)^{1/n}}, \label{eq:ser}
\end{equation} 

where,
 $\rho(r)$ is the density at radius $r$, $\rho_0$ is the central density. $R_e$ is the effective radius, which is the radius that encloses half of the total luminosity. $p$ is a parameter related to the Sérsic index $n$, typically $p = 1 - 1/(2n)$ for the deprojected Sérsic profile.
 $b$ is a constant that depends on the Sérsic index $n$ and is defined such that $R_e$ encloses half of the total luminosity. $n$ is the Sérsic index, which describes the concentration of the profile.

The parameter $b$ is usually approximated as $b \approx 2n - 1/3 + 0.009876/n$ for large values of $n$ \citep{ter05}. The Sérsic profile is a generalization of the de Vaucouleurs profile (for $n = 4$) and the exponential profile (for $n = 1$).
Below, we describe the structural parameters obtained from NIHAO for the sample of three galaxies and the fits we obtain using equations \ref{eq:dehn} and \ref{eq:ser}. The models exhibit axis ratios (b/a and c/a) as observed in reference snapshots for each component taken at a distance of 1 kpc from the center. Our $N-$body models have roughly 10 times higher mass resolution and 100 times higher spatial resolution when compared to their NIHAO counterparts.

\subsection{g3.78e13}
\begin{figure}
\centerline{
  \resizebox{0.98\hsize}{!}{\includegraphics{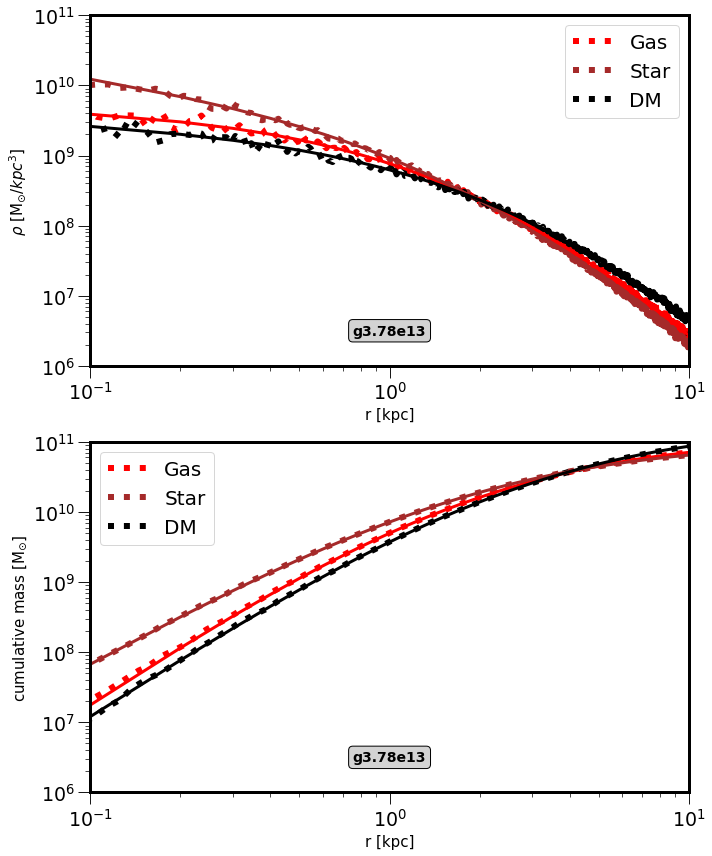}}
  }
%\centerline{
%  \resizebox{0.95\hsize}{!}{\includegraphics{figs/cum-masssini.eps}}
%  }
\caption{
Initial volume density profiles of gas, stars and dark matter components (top panel) and cumulative mass profiles of the same (bottom panel) for g3.78e13 galaxy shown as dashed lines. Dehnen model fits for parameters listed in table \ref{tab:table1} are shown as full lines in the same colour scheme.} \label{fig:g378ini}
\end{figure}

The gas, stellar and dark matter profiles for galaxy g3.78e13, centered on the density center of each component, are shown in Figure \ref{fig:g378ini}. Dashed lines represent profiles from snapshots taken from the NIHAO simulation, while solid lines are derived by fitting the \citet{deh93} profile. The galaxy and fit parameters are detailed in Table  \ref{tab:simparam}.

For g3.78e13, all components are well described by the Dehnen profile. The stellar mass and density dominate in the inner kiloparsec (kpc), while, the dark matter and gas components have comparable masses and densities within 1 kpc. The stellar mass exhibits a steeper profile ($\gamma$ = 0.6) compared to the gas ($\gamma$ = 0.1) and dark matter ($\gamma$ = 0.2). The mass fractions of gas, stars, and dark matter within the inner kpc are 0.31, 0.45, and 0.24, respectively.

%-------------------------------------------------------------------------%
%------------------------------------------------------------------------

We generated N-body models with a total of 3 million particles (1 million for each galaxy component) from the fitted profiles using AGAMA \citep{vasi19}. This corresponds to a stellar particle of mass $5 \times 10^4$ and dark matter particle with a mass of  $9 \times 10^4$.

\subsection{g9.61e12}

\begin{figure}
\centerline{
  \resizebox{0.98\hsize}{!}{\includegraphics{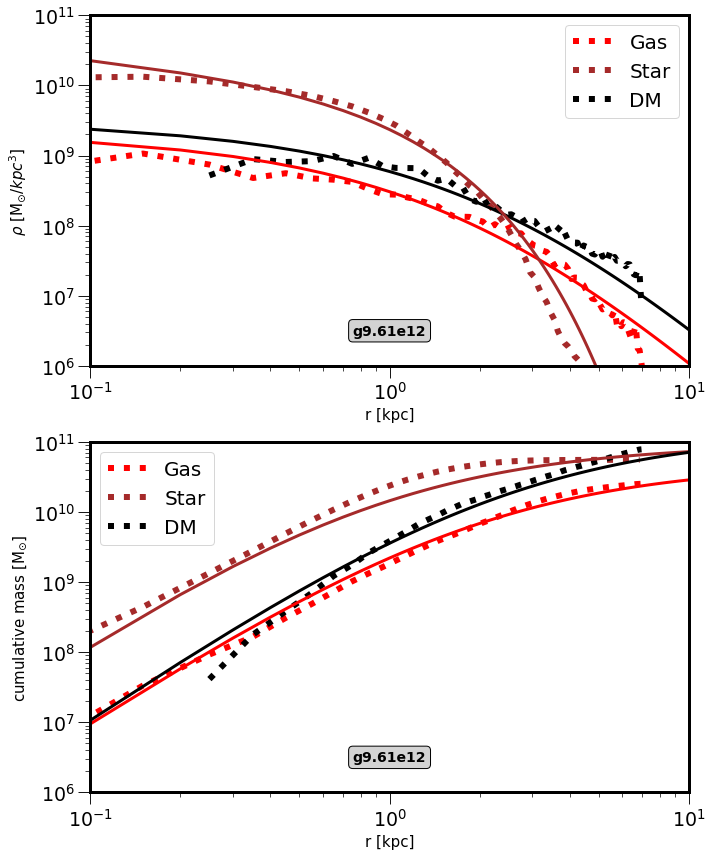}}
  }
%\centerline{
%  \resizebox{0.95\hsize}{!}{\includegraphics{figs/g961-cummasssini.eps}}
%  }
\caption{
Initial volume density profiles of gas, stars and dark matter components (top panel) and cumulative mass profiles of the same (bottom panel) for g9.61e12 galaxy shown as dashed lines. Dehnen/Sersic model fits for parameters listed in table \ref{tab:table1} are shown as full lines in the same colour scheme.} \label{fig:g961ini}
\end{figure}

Figure \ref{fig:g961ini} presents the mass profiles and fits for galaxy g9.61e12. Here again, gas and dark matter profiles are well described by Dehnen models but stellar density drops much steeply than $r^{-4}$ in outer parts and hence cannot be fitted with a Dehnen profile. Instead, we fit it with a Sersic profile and it is well fitted with the one having Sersic index of 0.9 very close to 1 which represents an exponential profile. The stellar profile dominates by almost an order of magnitude in the inner kpc. For g9.61e12, the mass fractions of gas, stellar and dark matter components inside inner kpc are, $0.07, 0.74, 0.19$ respectively. 
Again we generated N-body models with a total of 3 million particles from the fitting profiles using AGAMA.

%-------------------------------------------------------------------------%
%------------------------------------------------------------------------

\subsection{(g4.81e12)}

\begin{figure}
\centerline{
  \resizebox{0.98\hsize}{!}{\includegraphics{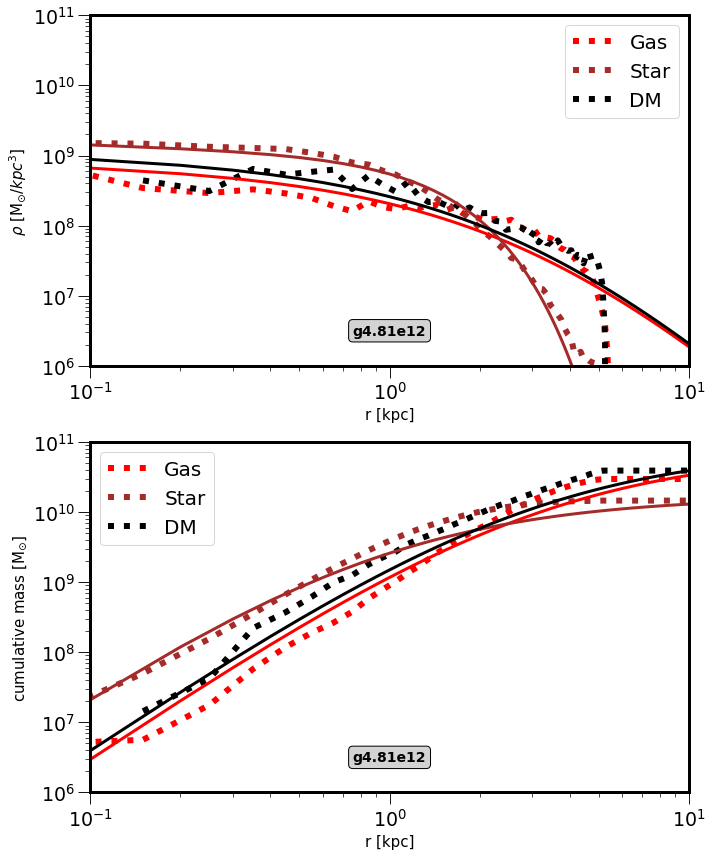}}
  }
%\centerline{
%  \resizebox{0.95\hsize}{!}{\includegraphics{figs/g81-cummasssini.eps}}
%  }
\caption{
Initial volume density profiles for stars and dark matter components (top panel) and cumulative mass profiles of the same (bottom panel) for g4.81e12 galaxy shown as dashed lines. Dehnen/Sersic model fits for parameters listed in table \ref{tab:table1} are shown as full lines in the same colour scheme.} \label{fig:g481ini}
\end{figure}

We plot the density and mass profiles for this model galaxy in figure \ref{fig:g481ini}. As earlier, dark matter and gas components are fitted well by Dehnen profiles but stellar profile is described by the Sersic profile with index 0.6. Stellar mass dominates in the very central 100 pc almost by an order of magnitude and becomes comparable to dark matter contents as we go beyond 1 kpc. Gas contributions around 1 kpc are 14 percent,  stellar mass is 52 percent whereas dm mass is about 34 percent of the total mass.

%-------------------------------------------------------------------------%
%------------------------------------------------------------------------

%N-body generated models and their stability is presented in figure \ref{fig:g481stabtri}. 

\section{$\phi$-GPU code} \label{sec:code}

$N$-body simulations presented in this section were performed using direct $N$-body code $\phi$-GPU \citep{berczik+11,Berczik2013} designed to run on massively parallel Graphical Processing Unit (GPU) supported clusters.  $\phi$-GPU calculates pairwise forces for all the active particles and uses fourth-order Hermite scheme with individual block time-step for orbit integration. The code uses Plummer-type gravitational softening for force calculations across all particle interactions, with each particle assigned its own individual softening length. Interaction softening between two particles is calculated using

\begin{equation}
    \epsilon_{ij} = \sqrt{\epsilon_{i} ^2 + \epsilon_{j} ^2}
\end{equation}

We have used $\phi$-GPU extensively for a wide range of astrophysical simulations including but not limited to SMBH dynamics in galaxy mergers \citep{khan+11,Khan+15,ber24} and IMBH dynamics in nuclear star clusters \citep{Khan_Holley-Bockelmann2021}.

First, we perform stability analysis runs to verify that our models are generated in dynamical equilibrium and that the density profiles and mass distributions of our model remain stable over the duration of our runs. The gas and stellar particles are assigned a softening length of \(\epsilon_{gas}\) and \(\epsilon_{\star}\) = 1 pc, respectively. On the other hand, the dark matter is given a much larger softening length, \(\epsilon_{dm}\) = 500 pc, to ensure it remains a collisionless component.  Details of the stability analysis are provided in Appendix A, where we confirm that the models maintain dynamical equilibrium throughout the simulation runs.

\section{Effect of Gas Removal} \label{sec:gasremoval}

In section \ref{NIHAO}, we have seen that for NIHAO simulations suit, SMBH feedback almost instantly (within few hundred Myr) removes almost all gas from the center of the galaxy causing gas surface density to drop by several orders of magnitudes. To replicate the process, we remove all gas from our N-body simulations and then evolve the remaining stellar and dark matter systems for 1 Gyr time using $\phi-$GPU for all our sample of galaxies. Again, stellar softening is chosen to be 1 pc and dark matter particles are given 500 pc to make the system collisionless. Below, we discuss the impact of gas removal on each of our three galaxies individually.

\subsection{g3.78e13} \label{g378runnogas-DM500pc}

\begin{figure}
\centerline{
  \resizebox{0.98\hsize}{!}{\includegraphics{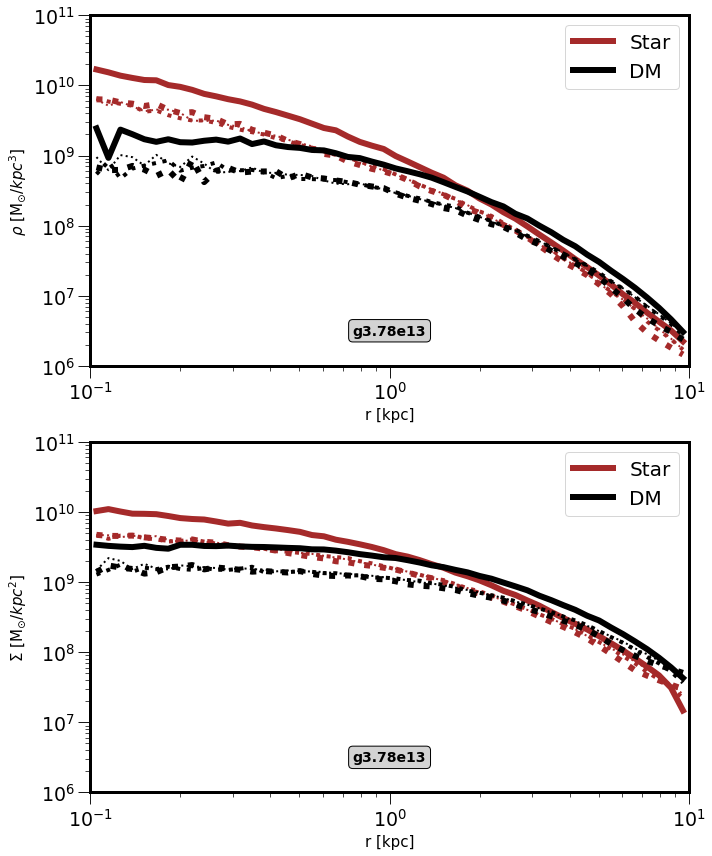}}
  }
%\centerline{
%  \resizebox{0.95\hsize}{!}{\includegraphics{figs/g378sigma1nogas-DM500pc.eps}}
%  }
\caption{
Time evolution of the volume density (top panel) and surface density (bottom panel) for stellar (brown lines) and dark matter profiles (black lines) in the absence of gas potential for g3.78e13 galaxy. Full lines are the initial unperturbed profiles while dashed lines represent the profiles at times of 100 Myr, 200 Myr and 400 Myr of evolution. Here, thicker lines represent earlier times and later times are represented by the thinner lines.}\label{fig:g378nogas500tri}
\end{figure}

As we remove gas from the system, both the stellar and dark matter distributions respond to the absence of gas potential and expand, resulting in drop in inner stellar and dark matter densities (figure \ref{fig:g378nogas500tri}). We plot the profiles for three different times (100 Myr, 200 Myr and 400 Myr) with dotted lines. As can be seen from the plots, all changes happen within first 100 Myr and at later times the profiles do not evolve much. The expansion or density drop seems to become more prominent as we move towards the center. 

%-------------------------------------------------------------------------%
%------------------------------------------------------------------------

\begin{figure}
\centerline{
  \resizebox{0.98\hsize}{!}{\includegraphics{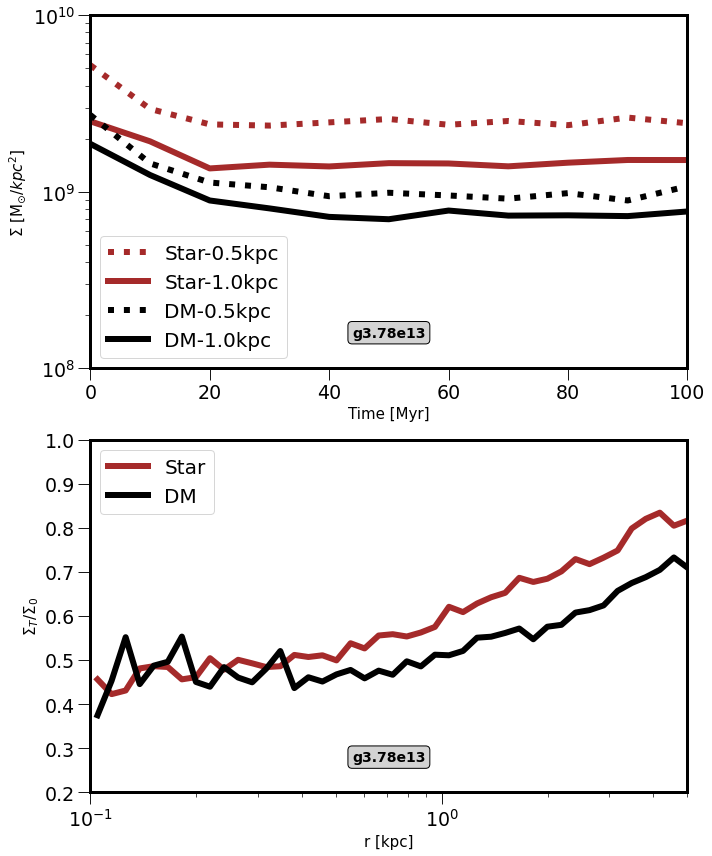}}
  }
%\centerline{
%  \resizebox{0.95\hsize}{!}{\includegraphics{figs/g378rhoratio-nogas.eps}}
%  }
\caption{
Top panel:Time evolution of surface density at 0.5 kpc (dashed lines) and 1.0 kpc (full lines lines) for stellar and dark matter distributions in the run without gas potential for g3.78e13 galaxy. Bottom panel: The ratio of the stellar and dark matter surface density at later time (T=500 Myr) to initial values.} \label{fig:g378nogas500triatkpc}
\end{figure}

We also plot surface density values (see figure \ref{fig:g378nogas500triatkpc}) as a function of time at 0.5 kpc and 1 kpc for initial 100 Myr of evolution, a period during which most of the expansion happened. The density dropped rapidly in initial 20 Myrs after which both the stellar and dark matter distributions achieved a new equilibrium state. The bottom panel of the figure shows the ratios of stellar surface densities at later times (500 Myr) and initial value for stellar and dark matter distributions at distance of 1.0 kpc. We notice that there is roughly a factor two drop in stellar and dark matter surface densities and that the change is more pronounced towards the center.

%-------------------------------------------------------------------------%
%------------------------------------------------------------------------
\subsubsection{Gas Reintroduction} \label{sec:gasaddition}

 It is worth exploring whether or not the stellar and dark matter systems return to their original configurations once they experience the potential of gas again. First, we estimate a reasonable timescale at which gas can reach the center after being expelled out of the system. We estimate free fall time for an object under the gravitational influence of our system density distribution. 

\subsubsection{ Free Fall Time Scale} \label{runnogas-DM500pcff}

We consider a snapshot corresponding to 500 Myr time evolution of the run without the gas potential. The free fall time can be estimated using;

\begin{equation}
    t_{\rm {ff}} = \sqrt{\left( \frac{3\pi}{32G\rho} \right)} \simeq 0.5427 \frac{1}{\sqrt{G\rho}}
\end{equation} \label{eq:free-fall time}

In our case, G=1 and at 5 kpc combined stellar and DM densities are of the order of $\rho \sim 10^{-4}$ in model units, which results in $t_{\rm{ff}}$ $\sim 50$ Myr. Also, we try average density; total mass enclosed inside 5 kpc divided by volume.
Total enclosed mass inside 5 kpc is 0.16+0.15 = 0.31 and volume is roughly $500$ kpc$^3 $ resulting in density to be ~ $\rho \sim 6.0 \times 10^{-4}$ and hence free fall time according to equation (1) is 23 Myr.
To confirm this estimate we also run a numerical experiment, where we drop in a free-falling particle at a distance of 5 kpc and use our code to integrate its motion. This direct experiment gave us a free-fall time of 26 Myr, in very good agreement with the theoretical estimation. 

%We also run a fuducial run where we try to estimate free-fall time of a test mass dropped at 5 kpc to compare with our theoretical calculations based on the standard free file time formula: Our choice of a massive particle is motivated by the fact that a particle with mass comparable to those of stellar or dark matter particles will undergo strong scatterings from 2-body interactions resulting in significant changes in its trajectory and free fall time as well.  From test MBH, we notice that it reaches center in about 26 Myr, %( figure (\ref{fig:g378bhff})%, so our estimate with the use of average density inside 5 kpc seems very reasonable. 

%-------------------------------------------------------------------------%
%\begin{figure}
%\centerline{
%  \resizebox{0.95\hsize}{!}{\includegraphics{figs/g378bhff.eps}}
%  }
%\caption{
%Separation of a test SMBH released with zero velocity at an initial separation of 5 kpc. %/Users/fmk5060/work/avm-projects/g378model1/g378-agamatriaxial/runnogas-DM500pcff/g378bhff.eps} %\label{fig:g378bhff}
%\end{figure}
%------------------------------------------------------------------------

\subsubsection{Gas External Potential} 

We choose snapshot at $t=500$ Myr of evolution (after gas removal), we recenter the system at the origin (including the SMBH) and then we (re)introduced gas potential as an external Dehnen potential with the exact same total mass and profile as it was calculated from the NIHAO snapshot.

\begin{figure}
\centerline{
  \resizebox{0.98\hsize}{!}{\includegraphics{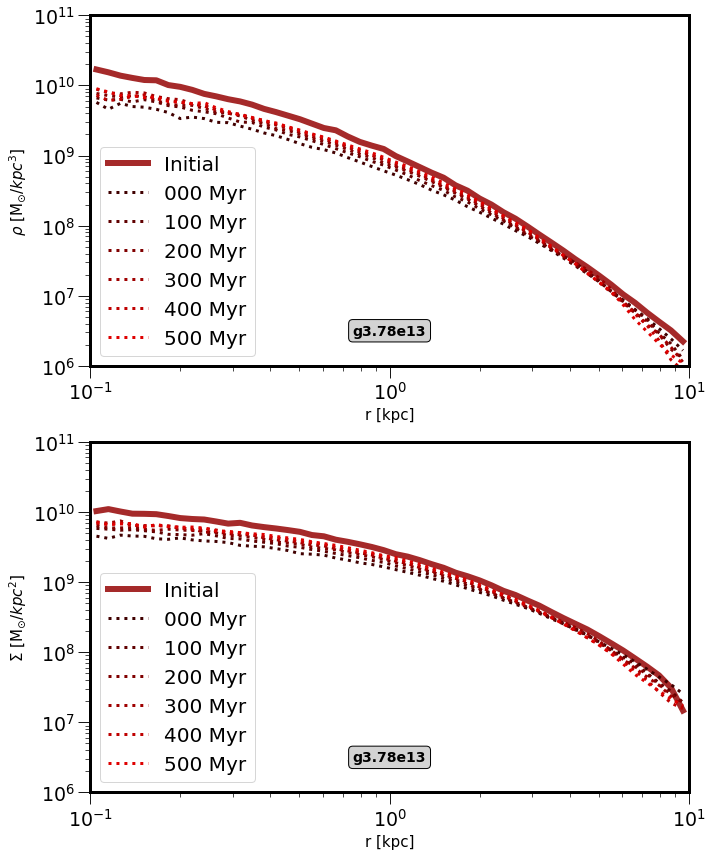}}
  }
%\centerline{
%  \resizebox{0.95\hsize}{!}{\includegraphics{figs/sigma1-starr10pc.eps}}
%  }
\caption{
Time evolution of volume density (top panel) and surface density profiles (bottom panel) for stellar distribution after reintroduction of gas component as an external potential in steps of 10 percent with time interval of 50 Myr. Full lines are initial unperturbed profile while dashed lines represent the profiles at various times of evolution till a time of 500 Myr.} \label{fig:g378gasext10pcstartri}
\end{figure}

\begin{figure}

\centerline{
 \resizebox{0.98\hsize}{!}{\includegraphics{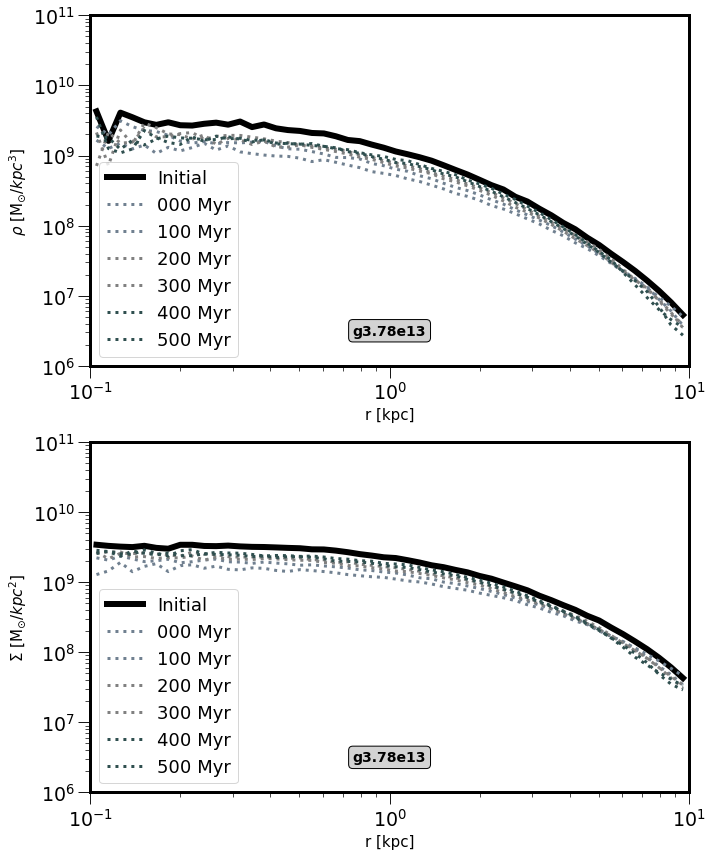}}
 }
%\centerline{
%\resizebox{0.95\hsize}{!}{\includegraphics{figs/sigma1-dm10pc.eps}}
%  }
\caption{
Time evolution of density (top panel) and surface density profiles (bottom panel) for dark matter after reintroduction of gas component as an external potential in steps of 10 percent with time interval of 50 Myr. Full lines are initial unperturbed profile while dashed lines represent the profiles at various times of evolution till a time of 500 Myr.} \label{fig:g378gasext10pcdmtri1}
\end{figure}

Although the gas was ejected rapidly due to AGN feedback, its re-accumulation in galaxy centers occurs on longer, cosmological timescales, as it must cool and radiate energy while traveling from the intergalactic medium and halo back to the center. To better replicate these timescales, we introduce the gas potential incrementally, using $10\%$ of the total gas mass over time intervals of 50 Myr, which is nearly double the free-fall time. The complete gas potential is reached after 450 Myr, with the simulation ending at $T = 500$ Myr. We notice that stellar and dark matter density increases as we add more mass (figures \ref{fig:g378gasext10pcstartri} and \ref{fig:g378gasext10pcdmtri1}), however it never reaches the original value (full lines).

%------------------------------------------------------------------------

%-------------------------------------------------------------------------%
%------------------------------------------------------------------------

%-------------------------------------------------------------------------%
\begin{figure}
\centerline{
  \resizebox{0.98\hsize}{!}{\includegraphics{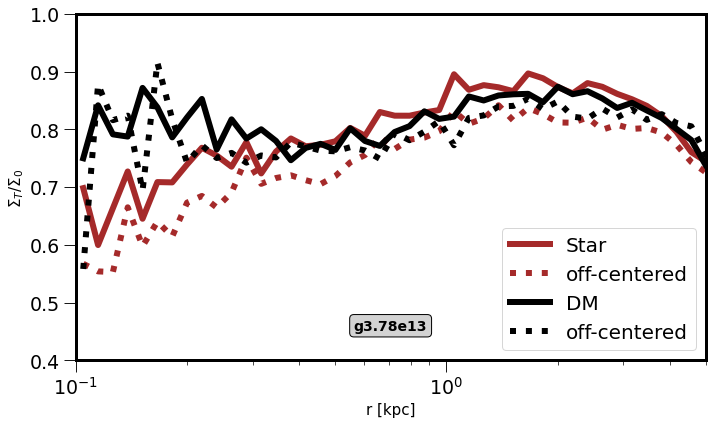}}
  }
\caption{
Ratio of final to initial surface densities for dark matter and stars as a function of the distance from the center.} \label{fig:g378ratio-gasext10pc}
\end{figure}
%------------------------------------------------------------------------

We plot the ratios of stellar and dark matter surface densities at the final time for this run to those of initial reference values (figure \ref{fig:g378ratio-gasext10pc}). The values witnessed here are approximately 25 percent smaller than reference value and the difference become smaller and smaller as we move away from the center to larger $r$. 

\subsubsection{Gas addition 10 percent off-center}

In NIHAO galaxy sample that we are modeling here, we noticed that gas distribution can be off-centered by a few hundred parsec to a kpc with respect to the dynamical center of the galaxy (dark matter and stellar combined).  We study the impact of an off centered gas distribution relative to stellar and dark matter distributions on central dark matter and stellar profiles. We moved the center of gas distribution off by 500 pc replicating the measured values in our reference snapshot of g3.78e12 galaxy. %Figures \ref{fig:g378gasoffext10pcst} and \ref{fig:g378gasoffext10pcdm} show how stellar and dark matter distributions evolve in the off centered gas potential. 
From figure \ref{fig:g378ratio-gasext10pc}, notice that ratio of surface densities (black dashed line) at the end of current run to that of initial value remains more or less same for dark matter profile but the change becomes more pronounced for the stellar component (brown dashed line) towards the center.

\subsection{g9.61e12} \label{subsec:galaxyg961}

\begin{figure}
\centerline{
  \resizebox{0.98\hsize}{!}{\includegraphics{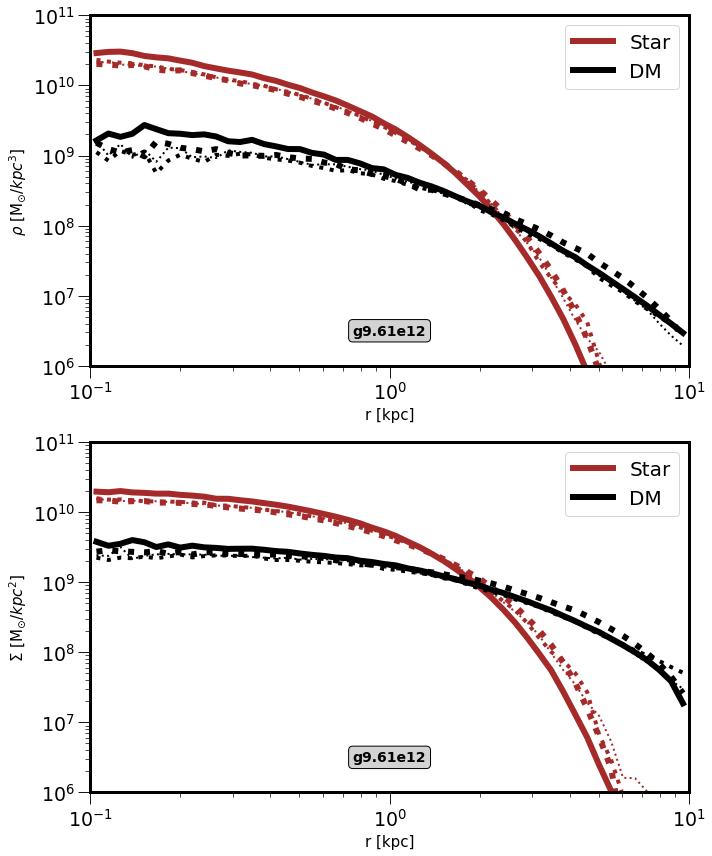}}
  }
%\centerline{
%  \resizebox{0.95\hsize}{!}{\includegraphics{figs/g961sigma1-nogas.eps}}
%  }
\caption{
Time evolution of the volume density (top panel) and surface density (bottom panel) for stellar (brown lines) and dark matter profiles (black lines) in the absence of gas potential for g9.61e12 galaxy. Full lines are the initial unperturbed profiles while dashed lines represent the profiles at times of 100 Myr, 200 Myr and 400 Myr of evolution.} \label{fig:g961nogas500tri}
\end{figure}

Here we present our results of the run as we remove the gas from the model for g9.61e12. Figure \ref{fig:g961nogas500tri} shows the evolution of stellar and dark matter profiles. As we noticed earlier that for g961 model the gas fraction is very small (0.07) relative to the total enclosed mass inside 1 kpc, we witness only a small change in original distributions in response to removal of gas potential.   

%-------------------------------------------------------------------------%
%------------------------------------------------------------------------

\begin{figure}
\centerline{
  \resizebox{0.98\hsize}{!}{\includegraphics{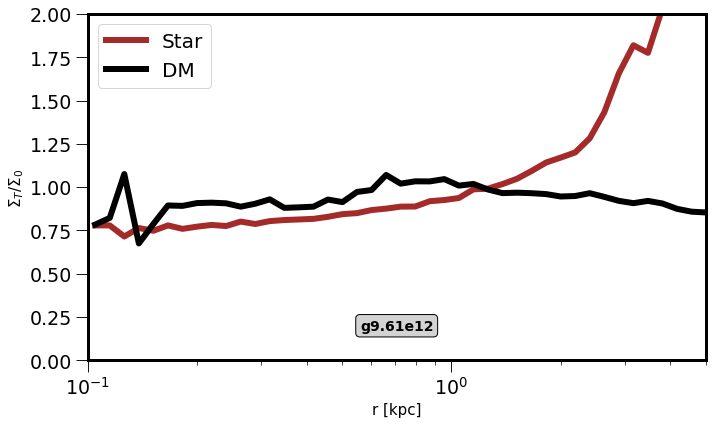}}
  }
  \centerline{
  \resizebox{0.98\hsize}{!}{\includegraphics{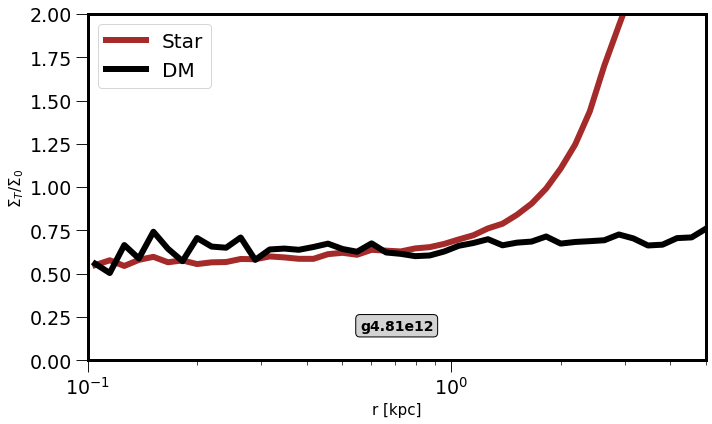}}
  }
\caption{
Ratio of final surface density at the end of gas expulsion runs to that of the initial unperturbed ones for galaxies g9.61e12 (top panel) and g4.81e12 (bottom panel) as a function of distance for stellar and dark matter profiles. } \label{fig:g961481rationogas}
\end{figure}

In order to have a more quantitative estimate of the central density profile changes caused by the gas expulsion, we plot the ratio of the final (at t=600 Myr) to initial density profiles for stellar and dark matter profiles (top panel of figure \ref{fig:g961481rationogas}). We notice that close to the center stellar distribution undergoes a decrease of roughly 25 percent and gradually increases as we move to larger distance. The mass removed from the central part is deposited at larger radii resulting in the increase of stellar density. We observe similar trends in the dark matter distribution in the inner regions. However, the increase in surface density at larger distances, seen in the stellar component, is absent for dark matter. This is due to the significantly larger dark matter mass already present at greater distances compared to the stellar mass.     

%\subsubsection{gas addition} \label{sec:g961gasplus}

%We reintroduce the gas mass as an external potential increasing linearly as a function of time reaching the full value in 500 Myr. Figure \ref{fig:g961gastt} presents time evolution of stellar and dark matter profiles which recover but not to the exact same value.

%-------------------------------------------------------------------------%
%------------------------------------------------------------------------

%-------------------------------------------------------------------------%
%\begin{figure}
%\centerline{
%  \resizebox{0.95\hsize}{!}{\includegraphics{figs/g961rho-gasextpott.eps}}
%  }
%\centerline{
%  \resizebox{0.95\hsize}{!}{\includegraphics{figs/g961cummass-gasextpott.eps}}
%  }
%\caption{
%Time evolution of density (top panel) and cumulative mass profile (bottom panel) for stellar and dark matter profiles as gas potential is restored proportionally with time reaching its full value in 500 Myr. Full lines are intial unperturbed profile while dashed lines represent the profiles at times of 100 Myr, 300 Myr and 600 Myr of evolution.} \label{fig:g961gastt}
%\end{figure}
%------------------------------------------------------------------------

\subsection{g4.81e12} \label{subsec:galaxyg4811}
\begin{figure}
\centerline{
  \resizebox{0.98\hsize}{!}{\includegraphics{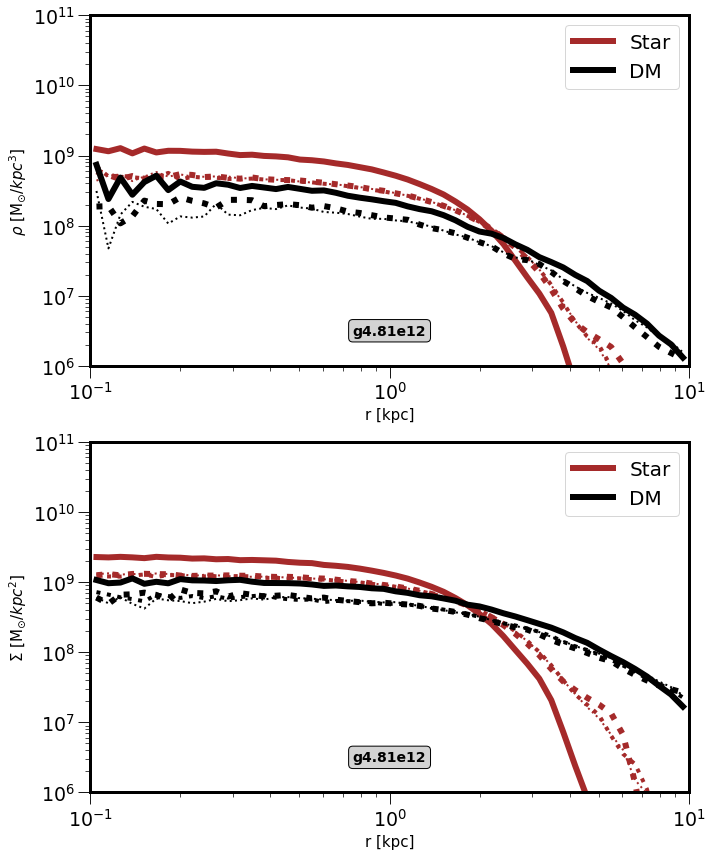}}
  }
%\centerline{
%  \resizebox{0.95\hsize}{!}{\includegraphics{figs/g481sigma1-nogas.eps}}
%  }
\caption{
Time evolution of the volume density (top panel) and surface density (bottom panel) for stellar (brown lines) and dark matter profiles (black lines) in the absence of gas potential for g4.81e12 galaxy. Full lines are the initial unperturbed profiles while dashed lines represent the profiles at times of 100 Myr, 200 Myr and 400 Myr of evolution. } \label{fig:g481nogas500tri}
\end{figure}

This is the least massive galaxy in our study sample of 3 galaxies for which we performed high resolution $N-$body simulations. As earlier, we again turn off external potential representing the gas component and study the system evolution for 500 Myr. The figure \ref{fig:g481nogas500tri} shows the evolution of volume densities for stellar and dark matter components in top panel whereas the bottom panel shows the surface densities for the same. 

%-------------------------------------------------------------------------%
%------------------------------------------------------------------------

Again we notice that that both stellar and dark matter density evolve very rapidly in first 100 Myr and remain stable subsequently.
%-------------------------------------------------------------------------%
%\begin{figure}
%\centerline{
%  \resizebox{0.98\hsize}{!}{\includegraphics{figs/agn-g481sigmas.png}}
%  }
%\caption{
%Ratio of final surface density at the end of gas expulsion run to that of the initial unperturbed one for galaxy %g4.81e12 as a function of distance for stellar and dark matter profiles. } \label{fig:g481rationogas}
%\end{figure}
%------------------------------------------------------------------------
Stellar surface density when compared to that of initial profile drops by almost a half in inner kilo parsec (bottom panel of figure \ref{fig:g961481rationogas}) as witnessed for the g3.78e13 case. In the outer parts, density increases due to expansion of the inner profile which shows that effect of gas removal is the expansion in inner part and outer parts remain unperturbed rather their density increases as mass flows outwards. 

%We reintroduce gas potential linearly till it reaches it's full value in 500 Myr. The resulting evolution of profiles is shown in figure \ref{fig:g481-gasextt}. Again we notice that profiles are unable to reach their initial value.

%-------------------------------------------------------------------------%
%\begin{figure}
%\centerline{
%  \resizebox{0.95\hsize}{!}{\includegraphics{figs/g481rho-gasextt.eps}}
%  }
%\centerline{
%  \resizebox{0.95\hsize}{!}{\includegraphics{figs/g481cummass-gasextt.eps}}
%  }
%\caption{
%Time evolution of density (top panel) and cumulative mass profile (bottom panel) for stellar and dark matter profiles as gas potential is restored  proportionally with time reaching its full value in 500 Myr. Full lines are intial unperturbed profile while dashed lines represent the profiles at times of 00 Myr, 300 Myr and 600 Myr of evolution. } \label{fig:g481-gasextt}
%\end{figure}
%------------------------------------------------------------------------

\section{Summary and Discussion} \label{sec:summary}

In this study, we used both the cosmological and controlled $N-$body simulations to analyse the evolution of a key scaling relation in elliptical galaxies, namely $\Sigma_1 - M_\star$ relation which relates the stellar surface density inside 1 kpc to the total stellar mass of the galaxy. The first part of this work presents the analysis of a sample of galaxies obtained from NIHAO cosmological simulation suite \citep{Wang2014}. We look how $\Sigma_1 - M_\star$ looks at redshift 0 for our selected sample of galaxies and also how it evolves for individual galaxies during their time evolution. Subsequently, we run direct $N-$body simulations to validate some of our findings. We list key findings from the study as follows:

(i) NIHAO galaxies with AGN feedback follow the observed $\Sigma_1 - M_\star$ relation reasonably well all the way to the observed flattening for high mass galaxies at redshift zero. On the other hand for galaxies without AGN feedback $\Sigma_1$ continues to increase for high mass galaxies contradicting the observed flattening for the relation.

(ii) The trajectory of individual galaxies as a function of time in the $\Sigma_1 - M_\star$ plane follows a single power law for both galaxies with and without AGN feedback till a stellar mass of about few $10^{10} M_{\odot}$ is reached. At higher masses galaxies without AGN continue to move along a single power-law, while galaxies with AGN feedback have first a declining $\Sigma_1$ and then they stabilize around a constant value for the inner density.

%during their cosmic evolution goes hand in hand for both the cases, with and without AGN feedback till the galaxies reach a mass of about $M_{\star} \simeq 10^{11} M_{\odot}$ beyond which $\Sigma_1$ flattens for the galaxies with AGN feedback and follows the near linear slope for those with no AGN feedback.

(iii) This transition mass coincides with the peak of the SMBH accretion, and the subsequent expulsion of large quantities of gas from the inner region, which causes the stellar (and dark matter) density to react via a non-adiabatic expansion. This stellar expansion causes a first decrease of $\Sigma_1$, which then is kept constant for the last part of the evolution due to the lack of star formation due to the gas removal. 

(iv) Our results suggest that the AGN activity has a key role in first reducing and then regulating the amount of stellar mass in the center of massive galaxies.

%(iv) It also coincides with the ejection of gas from the center caused by the feedback from SMBH during AGN phase. The gas expulsion happens at relatively shorter timescales ~ few hundred Myr, resulting in sudden drop of gas mass by approximately two orders of magnitude. 

%(v) The ejection of gas depletes the central gas reservoir, leading to a reduction in SMBH accretion and halting its mass growth. This also suppresses central star formation, which in turn causes the stagnation of $\Sigma_1$.

%(vi)The sudden drop in central gas content in galaxies with AGN feedback is accompanied by a nearly twofold decrease in $\Sigma_1$. This reduction in gas mass leads to a shallower gravitational potential, prompting the stellar and dark matter distributions to react and readjust accordingly.

(v) To confirm our results based on cosmological simulations, we run a set of controlled $N-$body experiments (at much higher resolution) based on physical parameters obtained from NIHAO galaxies. We confirm that the reduction of $\Sigma_1$ happens roughly on the dynamical timescale and both the stellar and dark matter distributions remain stable beyond that.

(vi) Finally we explore a scenario where the galaxy is able to reassemble gas in its central regions on free fall timescales, and we show that even if the galaxy is able to gain back the same amount of gas that was lost via AGN feedback, the stellar (and dark matter) density will never fully return to their pre-expansion values.

%n Although the galaxies never fully regain the gas mass present before peak AGN activity, we conducted several runs to study the impact of gradual gas reassembly in the center, modeled as a linear growth over 500 Myr. We observed that the stellar and dark matter profiles recover by about 50\%, but they never fully return to their pre-expansion values.

Our study demonstrates that AGN feedback, and the associated expulsion of gas driven by the immense energy released during this phase, plays a pivotal role in flattening the $\Sigma_1 - M_\star$ relation for massive galaxies with stellar masses of $M_{\star} \simeq 10^{11} M_{\odot}$. As already mentioned (see introduction) there are other possible mechanism that can act together with feedback induced expansion to shape the central distribution of stars in massive galaxies, as for example core scouring by SMBH binaries \citep{graham+04,mer06,khan+12a} and dynamical effects of SMBH  recoil \citep{2008ApJ...678..780G,Khonji:2024}; these two other channels have recently been shown to be quite important in the context of the evolution of the so-called Little Red Dots \citep{Khan2025}.

On the other hand, due to the resolution of our cosmological simulations and the absence of SMBH mergers in our numerical experiments, our results show that, in principle, feedback induced expansion can significantly account for observed $\Sigma_1 - M_\star$ relation for massive galaxies. We defer to future studies the investigation of the relative impact of different mechanisms in altering the stellar distribution in massive objects.

Finally, we would like to underline that our results can provide  one of the first observational evidence for the so-called non-adiabatic expansion of a collisionless component, a mechanism that has been invoked to alleviate most (if not all) the problems of cold dark matter on the scale of dwarf galaxies \citep{Pontzen2012,Dutton2016, Bullock2017}.

%Below this threshold, the supermassive black hole is typically less massive, and the central gas reservoir is still accumulating—both prerequisites for sustaining the strong AGN feedback necessary to drive significant gas expulsion and influence the central stellar density profile.

\section*{Acknowledgments}

 The authors are grateful to Nik Arora for useful discussion at the early stages of this work. The authors acknowledge support from High Performance Computing resources at New York University Abu Dhabi. This material is based upon work supported by Tamkeen under the NYU Abu Dhabi Research Institute grant CASS.
%\section*{Data Availability Statement}
%The data underlying this article will be shared on reasonable request to the corresponding author.

\bibliography{ms}{}
\bibliographystyle{aasjournal}

%% This command is needed to show the entire author+affiliation list when
%% the collaboration and author truncation commands are used.  It has to
%% go at the end of the manuscript.
%\allauthors

%% Include this line if you are using the \added, \replaced, \deleted
%% commands to see a summary list of all changes at the end of the article.
%\listofchanges

\newpage
\appendix
\section{Stability Analysis}

\begin{figure*}[h!]
\centering{
  \resizebox{0.76\hsize}{!}{\includegraphics{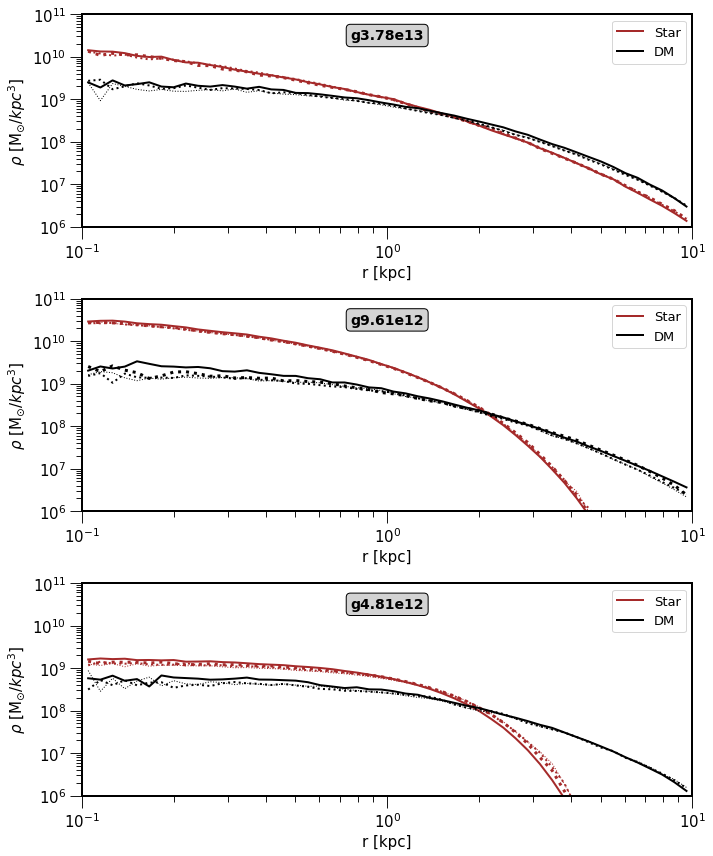}}
  }
\caption{Volume density profiles of the individual components for various galaxies. The full lines represent the initial profiles and the dotted lines represent the model's evolution over 1 Gyr in N-body simulations. The close match between solid and dotted lines indicates stability throughout the run.}
    \label{fig:stab_all}
\end{figure*}

We evolve all our N-body models for 1 Gyr using $\phi$-GPU to check for their stability. Figure \ref{fig:stab_all} shows the time evolution of density and cumulative mass profiles of individual components for various times up to 1 Gyr (dotted lines). We can clearly see that profiles are practically indistinguishable from initial ones shown as full lines. Dark matter shows small evolution inside 1 kpc because of the large softening ($\epsilon_{dm} = 500 pc$) employed. As is usually the case with $N-$body codes, the softening effects diminish beyond 1-2 softening lengths, and we do not notice any difference in DM profiles beyond 1 kpc.

%-----------------------------------------------------------------------

\section{$\Sigma_1 - M_\star$ for selected galaxies}

Here we present $\Sigma_1 - M_\star$ for 8 NIHAO galaxies in addition to 6 presented in figure \ref{fig:nihaosigmavsmstar}. We witness that generally $\Sigma_1$ evolves while staying close to observed linear relation and once they reach a mass of $M_\star > 10^{10} M_{\odot}$, they grow above the flattened observed line and eventually fall back more in line with observations due to AGN feedback assisted expansion in the central region.

\begin{figure*}[h!]
\centering{
  \resizebox{0.94\hsize}{!}{\includegraphics{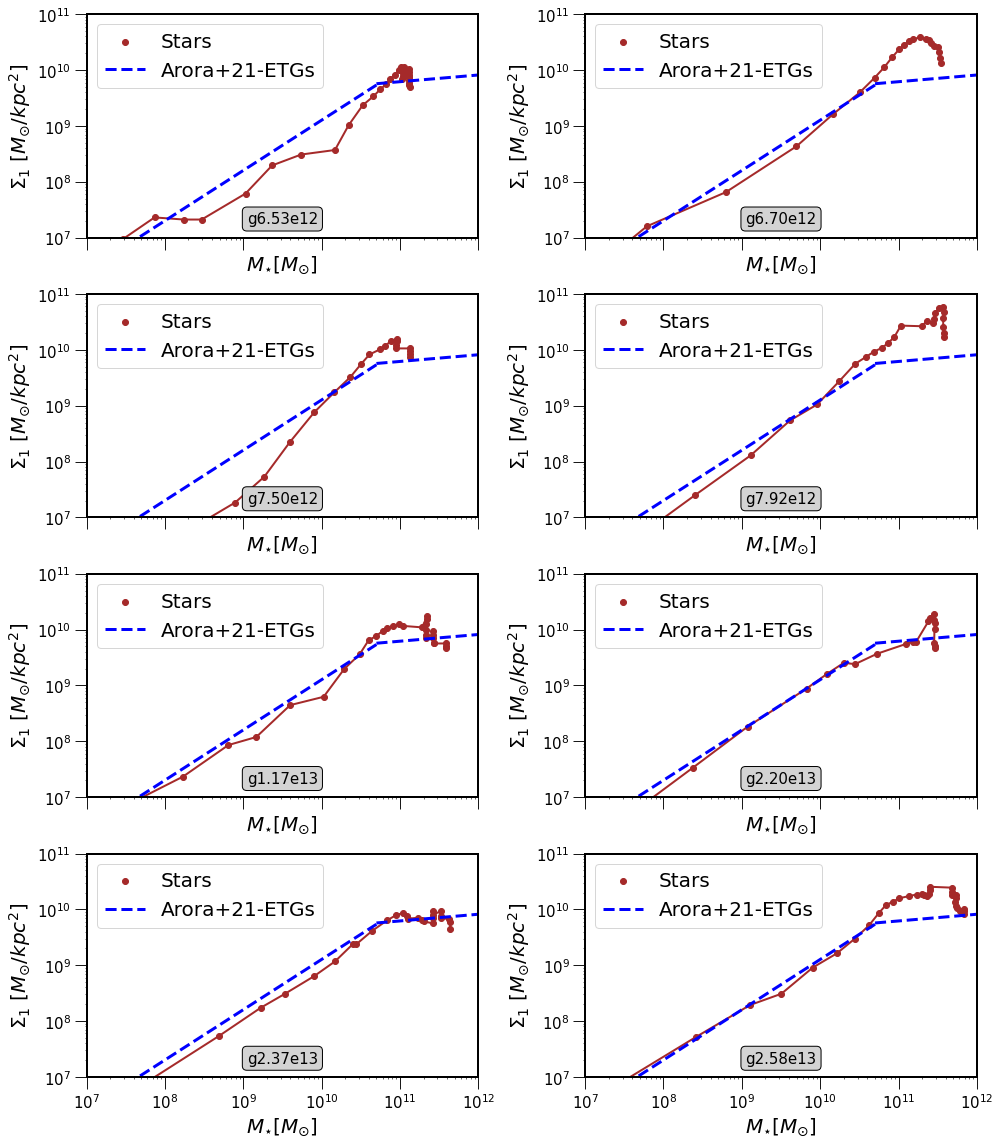}}
  }
\caption{$\Sigma_1 - M_\star$ for a sample of 8 high mass galaxies.}

% Again dashed blue line is piecewise fit form \citet{Arora21}}.
    \label{fig:sigmasappendix}
\end{figure*} 

% Again dashed blue line is piecewise fit form \citet{Arora21}}.
%-----------------------------------------------------------------------

\end{document}